\documentclass[floatfix,reprint,nofootinbib,amsmath,amssymb,epsfig,pre,floats,letterpaper,groupedaffiliation]{revtex4}

\usepackage{amsmath}
\usepackage{amssymb}
\usepackage{graphicx}
\usepackage{epstopdf}
\usepackage{hyperref}
\usepackage{dcolumn}
\usepackage{bm}
\usepackage{listings}
\usepackage{xcolor}
\usepackage{inconsolata}

\newcommand{\beq}{\begin{equation}}
\newcommand{\eeq}{\end{equation}}

\begin{document}

\title{False shares in verifiable secret sharing with finite field commitments}

\author{Hua Lu}
\author{Jack Peterson}
\email{jack@augur.net}
\affiliation{Forecast Foundation, San Francisco, CA 94107 USA}

\begin{abstract}
Verifiable secret sharing (VSS) is designed to allow parties to collaborate to keep secrets.  We describe here a method of fabricating false secret shares that appear to other parties to be legitimate, which can prevent assembly of the decryption key.  This vulnerability affects VSS schemes using verification commitments bounded to a finite field.
\end{abstract}

\maketitle

Verifiable secret sharing (VSS) schemes~\cite{Shamir_1979,Blakley_1979,Feldman_1987} rely on the assumption that parties can not reliably fabricate false secret shares which pass the verification process.  Here, we show that, for certain VSS implementations that use verification commitments bounded to a finite field, this assumption is incorrect.

Let there be $n$ shares of a secret and a threshold of $t$ secret shares required to get the secret. Choose a finite field $\mathbb{Z}_p$ and a generator $g \in \mathbb{Z}_p$. Each party associates with a unique non-zero identity $i \in \mathbb{Z}_p$ and creates a secret random polynomial,
\beq
P_i\left(z\right) = a_{i,0} + a_{i,1} z + \cdots + a_{i,{t-1}} z^{t-1},
\eeq
with coefficients $a_{i,j} \in \mathbb{Z}_p$, and decryption key given by $\sum_i{a_{i,0}}$.  Each party also computes \emph{verification commitments} ($c_{i,j}$),
\beq
c_{i,j} \equiv g^{a_{i,j}} \mod p,
\eeq
which are made available to all parties.

Next, each party computes and sends the $k^{\mathrm{th}}$ secret share, $P(k)$, to party $k$. Party $k$ checks the incoming share against the sending party's verification commitments:
\beq
g^{P_i(k)} = g^{\sum_{j=0}^{t-1}{a_{i,j} k^j}} = \prod_{j=0}^{t-1}{g^{a_{i,j} k^j}} = \prod_{j=0}^{t-1}{{c_{i,j}}^{k^j}}.
\eeq
Party $k$ assumes the share received from party $i$ is legitimate if it matches the verification commitments (available to all parties).

We now describe the vulnerability.  First, note the following:
\begin{enumerate}
\item The polynomial interpolation for recovering decryption key parts must occur in the field $\mathbb{Z}_p$.
\item The multiplicative order of $g$ divides $p-1$.
\end{enumerate}
Suppose one of the parties is an adversary. The adversary creates polynomial $P_i$ and posts verification commitments $c_{i,j}$. The adversary sends false secret shares $Q_k \neq P_i(k)$ such that
\beq
Q_k = P_i(k) \pmod{p-1}.
\eeq
The false shares check out, $g^{Q_i} = g^{P_i(k)}$; however, attempts to recover $P_i(0)$ using $Q_k$ will most likely fail. The adversary can now prevent the decryption key from being assembled simply by not sharing $P_i(0)$.

In principle, this problem is avoidable by using verification commitments which are not bounded to the finite field. For example, we are free to choose commitments $c_{i,j} \equiv g^{a_{i,j}}$.  However, note that $a_{i,j}$ is then on the order of $p$.  Since this is at least $1024$ bits, there is no practical way to store these verification commitments.
\\\\
\emph{Financial support for this work was provided by J.~Costello.}

\bibliographystyle{unsrt}
\bibliography{vss.bib}

\end{document}